\documentclass[aps,prx,twocolumn,superscriptaddress,letterpaper,longbibliography]{revtex4}
\usepackage{amssymb}
\usepackage{amsmath}
\usepackage{upgreek}
\usepackage{graphicx}
\usepackage{color}

\definecolor{darkblue}{rgb}{0,0,0.5}
\definecolor{lila}{rgb}{0.3,0,0.3}
\definecolor{turq}{rgb}{0,0.1,0.4}
\definecolor{lightblue}{rgb}{0.7,0.7,0.9}
\usepackage{url} 
\usepackage[pdftex,
 colorlinks=true,
 backref=page,
 linkcolor=darkblue, 
 filecolor=red,
 citecolor=turq, 
 urlcolor=lila, 
 pdftitle={Unbiased All-Optical Random-Number Generator},
 pdfauthor={Tobias Steinle and Johannes N. Greiner and Jorg Wrachtrup and Harald Giessen and Ilja Gerhardt},
 pdfsubject={},
 pdfkeywords={},
 pdfpagelabels=true, 
 breaklinks=false,
 plainpages=false,
 backref=false,
 bookmarks,
 bookmarksnumbered=true]{hyperref}

\newcommand{\done}{\ensuremath{\mathbf{1}}}
\newcommand{\dzero}{\ensuremath{\mathbf{0}}}

\makeatletter
\renewcommand*{\@fnsymbol}[1]{\ensuremath{\ifcase#1\or \dagger\or *\or \ddagger\or
		\mathsection\or \mathparagraph\or \|\or **\or \dagger\dagger
		\or \ddagger\ddagger \else\@ctrerr\fi}}
\makeatother
\begin{document}

\title{Unbiased All-Optical Random-Number Generator}

\author{Tobias Steinle}
\thanks{These two authors contributed equally}
\affiliation{University of Stuttgart, 4th Physics Institute, Pfaffenwaldring 57, 70569 Stuttgart, Germany}
\affiliation{present address: ICFO -- Institut de Ciencies Fotoniques, The Barcelona Institute of Science and Technology, 08860 Castelldefels, Barcelona, Spain}
\author{Johannes N.\ Greiner}
\thanks{These two authors contributed equally}
\affiliation{3. Institute of Physics, University of Stuttgart and Institute for Quantum Science and Technology, IQST, Pfaffenwaldring 57, D-70569 Stuttgart, Germany}
\author{J\"org Wrachtrup}
\affiliation{3. Institute of Physics, University of Stuttgart and Institute for Quantum Science and Technology, IQST, Pfaffenwaldring 57, D-70569 Stuttgart, Germany}
\affiliation{Max Planck Institute for Solid State Research, Heisenbergstra\ss e 1, D-70569 Stuttgart, Germany}
\author{Harald Giessen}
\affiliation{University of Stuttgart, 4th Physics Institute, Pfaffenwaldring 57, 70569 Stuttgart, Germany}
\author{Ilja Gerhardt}
\email{i.gerhardt@fkf.mpg.de}
\affiliation{3. Institute of Physics, University of Stuttgart and Institute for Quantum Science and Technology, IQST, Pfaffenwaldring 57, D-70569 Stuttgart, Germany}
\affiliation{Max Planck Institute for Solid State Research, Heisenbergstra\ss e 1, D-70569 Stuttgart, Germany}

\begin{abstract}
The generation of random bits is of enormous importance in modern information science. Cryptographic security is based on random numbers which require a physical process for their generation. This is commonly performed by hardware random number generators. These exhibit often a number of problems, namely experimental bias, memory in the system, and other technical subtleties, which reduce the reliability in the entropy estimation. Further, the generated outcome has to be post-processed to ``iron out'' such spurious effects. Here, we present a purely optical randomness generator, based on the bi-stable output of an optical parametric oscillator. Detector noise plays no role and post-processing is reduced to a minimum. Upon entering the bi-stable regime, initially the resulting output phase depends on vacuum fluctuations. Later, the phase is rigidly locked and can be well determined versus a pulse train, which is derived from the pump laser. This delivers an ambiguity-free output, which is reliably detected and associated with a binary outcome. The resulting random bit stream resembles a perfect coin toss and passes all relevant randomness measures. The random nature of the generated binary outcome is furthermore confirmed by an analysis of resulting conditional entropies.
\end{abstract}
\maketitle

\section{Introduction}
Random numbers are of utter importance in our everyday life, even if many of us are not into gambling or statistics~\cite{galton_n_1890}. The most crucial use of random numbers is strong cryptography -- securing modern communication, money transfers, and storage of sensitive information. The encryption keys which are used to unlock encrypted data are secured by mathematical hard problems, most notably the discrete logarithm problem or prime-number factorization. The underlying keys are based on random numbers. As recently shown, one of the most efficient attack vectors on modern cryptography is the supply of weak random numbers~\cite{lenstra__2012,becker__2013}, reducing the key space to a fraction of the mathematical probable: Assuming a modern encryption key with $N$ bits results in a key space of $2^N$ possibilities -- with large $N$ this requires a long time for a brute-force decryption process. When such a key is only based on $n\ll N$ possible outcomes of a random number generator, the decryption of the data might be a question of seconds.

\begin{figure}[hb]
  \includegraphics[width=\columnwidth]{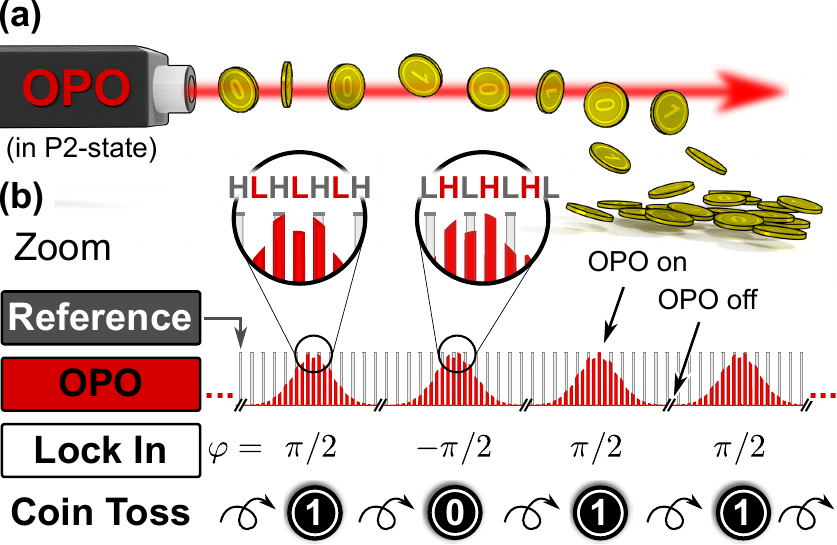}
  \caption{\textbf{Operation principle of the all-optical randomness generator.} a) The output of an optical parametric oscillator (OPO) generates two different output states unambiguously. Both outputs are equi-energetic and equi-probable, and are based on the transient oscillation of the OPO. We associate the outcomes to an output bit, comparable to a coin-toss. b) The detection is performed by a phase measurement ($\varphi$) against an external reference clock, supplied by the pump laser. \textbf{H} and \textbf{L} denote the different pulse energy outputs of the OPO, which operates in the period-doubling-state, named P2.}
  \label{fig:fig01}
\end{figure}

In the computer age, the first idea which might come to mind is a computer based randomness generator. Unfortunately, such generators are commonly defined based on a recurrence relation, and can only emit (partially very long) cycles of seemingly random bits~\cite{knuth__1968,bauke_josp_2004-02-01}. Therefore, hardware-based random number generators were presented in the past. The early hardware random number generators were a die~\cite{galton_n_1890} or simply a coin~\cite{feller__1968}. Both generators are well known even to non-scientists. In mathematical terms, a coin toss is a Bernoulli trial of the sample space $\Omega=\{\dzero,\done\}$ -- at least when the coin is not landing on its edge~\cite{murray_pre_1993}. A \emph{fair} coin is defined as a model system which exhibits no bias, cannot land on the edge, has no memory, and exhibits the probability $p(\dzero)=p(\done)=1/2$. This system is well-covered in literature~\cite{feller__1968,finkelstein_fq_1978,ford_pt_1983,vuloviifmmodeacutecelsecfi_pra_1986}. Besides classical random bit generators, which have to fulfill a number of requirements~\cite{stipcevic_rsi_2004}, a recent development are quantum-based generators, which utilize the inherently unpredictable nature of quantum effects to deliver random numbers~\cite{stefanov_jmo_2000,jennewein_rosi_2000,stipcevic_rsi_2007,furst_oe_2010,colbeck_phd_2009,pironio_n_2010,sanguinetti_prx_2014,jofre_true_2011}. 


\begin{figure*}[th]
  \includegraphics[width=\textwidth]{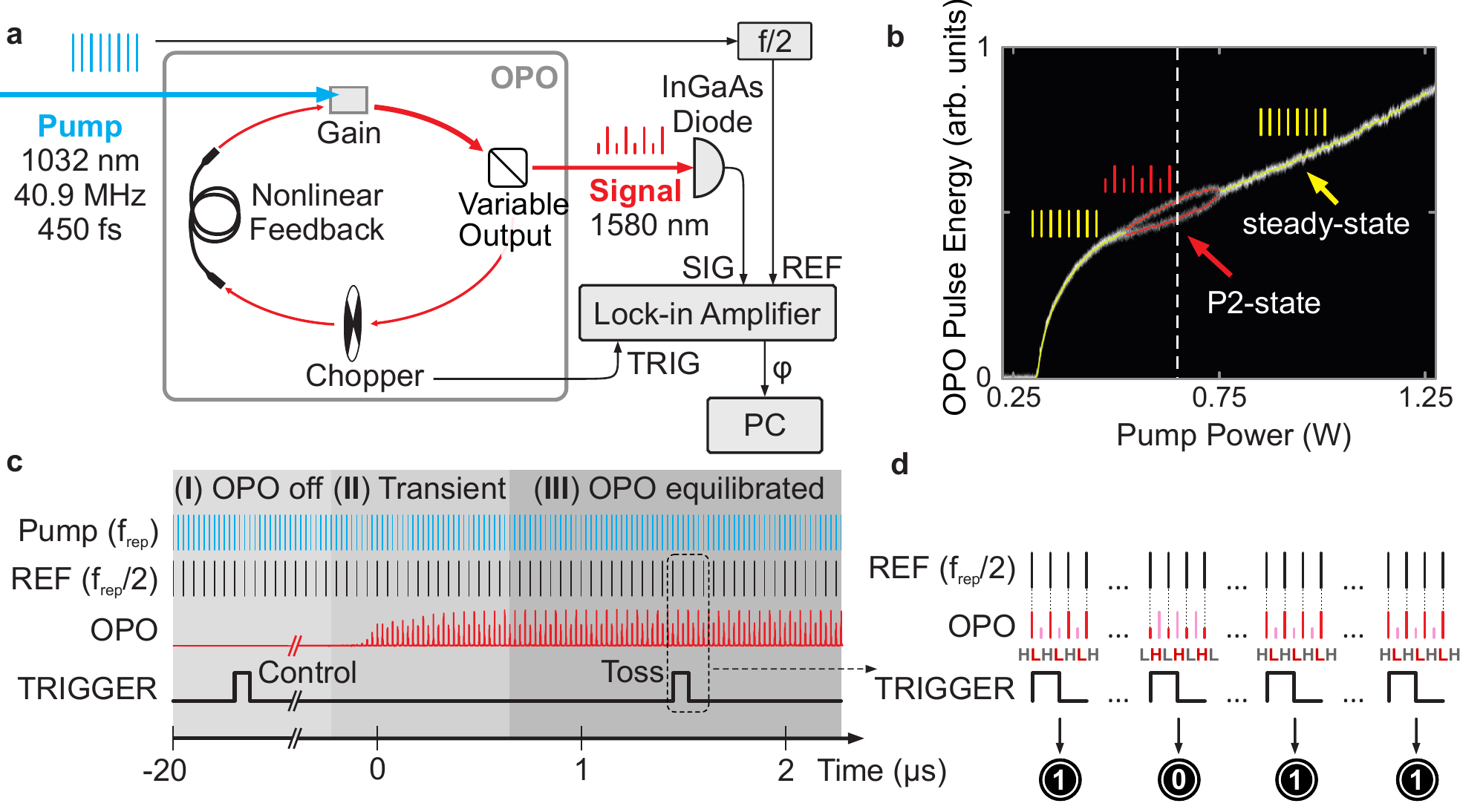}
  \caption{\textbf{Experimental scheme of randomness generation.} a) Experimental implementation of the optical parametric oscillator. b) Power dependent output pulse energy. To note that both different output pulse train options are equi-energetic. c) Measured transition scheme, periodic with the chopper frequency. The trigger pulse defines two measurements: One when the OPO is blocked (control), and one when the P2-regime is reached (toss). The reference frequency is 40.9~MHz/2, supplied by the pump laser and a frequency divider. d) Interpretation of measurement outcomes as final bits.}
  \label{fig:fig02}
\end{figure*}

For future applications, electrical circuits may eventually be completely replaced by solely optical devices due to the practical advantages of photons in terms of speed, leakage, heat development and wiring. Therefore, we introduce an ``all-optical'' randomness generation, in which the random process is independent of a particular detector implementation. A specific example are optical parametric oscillators (OPOs), in particular degenerate ones, which were used for this task before~\cite{marandi_oe_2012a,okawachi_ol_2016,marandi_oe_2012}. The relative phase of two generators results in a two-state outcome -- but it requires experimental efforts, such as two phase stabilized OPOs. As outlined in the literature, the OPO's outcoming phase is based on quantum processes, such that this represents another form of quantum randomness generation~\cite{herrero-collantes_rmp_2017,wang_coherent_2013,marandi_oe_2012,drummond_critical_2002,nabors_coherence_1990,harris_observation_1967,louisell_quantum_1961}. The generation of random numbers by an OPO has some advantages; these are the speed of an optical generator, its equi-energetic bi-stability, as well as a demodulator-based and ambiguity-free measurement principle. By ``ambiguity-free'' we refer to a measurement which has two (or more) definite outcomes, which can not be confused due to technical issues of the measurement apparatus. In quantum randomness generation with single photon detectors such ambiguities can occur for example due to dead-times, electrical jitter, and varying detection efficiencies~\cite{oberreiter_light_2016}.

Here we present the use of a bi-stable configuration implemented in a period-doubling optical parametric oscillator for randomness generation. To the best of our knowledge, this is the first experimental utilization of a P2-state in an OPO reported in the literature to date. A simplified model is depicted in Fig.~\ref{fig:fig01}. The involved bi-stability is equi-energetic and equi-probable; only two outcomes are possible and no bias is observed. For randomness generation, the stream of binary outcomes can be used directly, and no additional un-biasing or bit-extraction is required. We test the outcome against the predicted outcomes of a fair coin toss. At the end of the paper, we compute the most conservative bound, the min-entropy, against the size of a finite sample of bits originating from the generator.

\section{Experimental Scheme}

A home-built fiber-feedback optical parametric oscillator (OPO)~\cite{suedmeyer_ol_2001,steinle_oe_2015} is pumped by a mode-locked 450~fs, 1032~nm Yb:KGW oscillator (Fig.~\ref{fig:fig02}a). The gain element is a periodically poled lithium niobate crystal (PPLN). The repetition rate is defined by the laser and amounts to 40.9~MHz; the length of the OPO cavity is matched to this by a movable mirror. A part of the OPO cavity consists of a single-mode feedback fiber, which in combination with the variable output coupler allows to control the effective intracavity non-linearity. The output signal is detected on a reverse-biased InGaAs photo diode (Hamamatsu). The signal is monitored in real time on an oscilloscope (see Fig.~\ref{fig:fig02}c). Alternatively, the signal is fed into a lock-in amplifier for further analysis.

When the pump power is varied, the OPO exhibits a bi-modal behavior, which can be identified as period doubling~\cite{ikeda_oc_1979,lugiato_incd_1988,richy_jb_1995,steinmeyer_oc_1994,kues_oe_2009}. Above its oscillation threshold, the OPO operates in the steady-state (yellow trace in Fig.~\ref{fig:fig02}b), which results in an output pulse train with identical subsequent pulses, as known from any mode-locked laser. Upon further increase of pump power, the system enters the so-called period-2-state (P2-state) which delivers alternating pulses with different pulse energy, peak power, and spectral properties. This behavior originates from the interplay of spectral selective gain and nonlinear feedback~\cite{akhmediev_pre_2001}. As a result of the synchronous pumping of the OPO, these pulses are temporally aligned with the pump frequency.

\begin{figure}[ht!]
  \includegraphics[width=0.9\columnwidth]{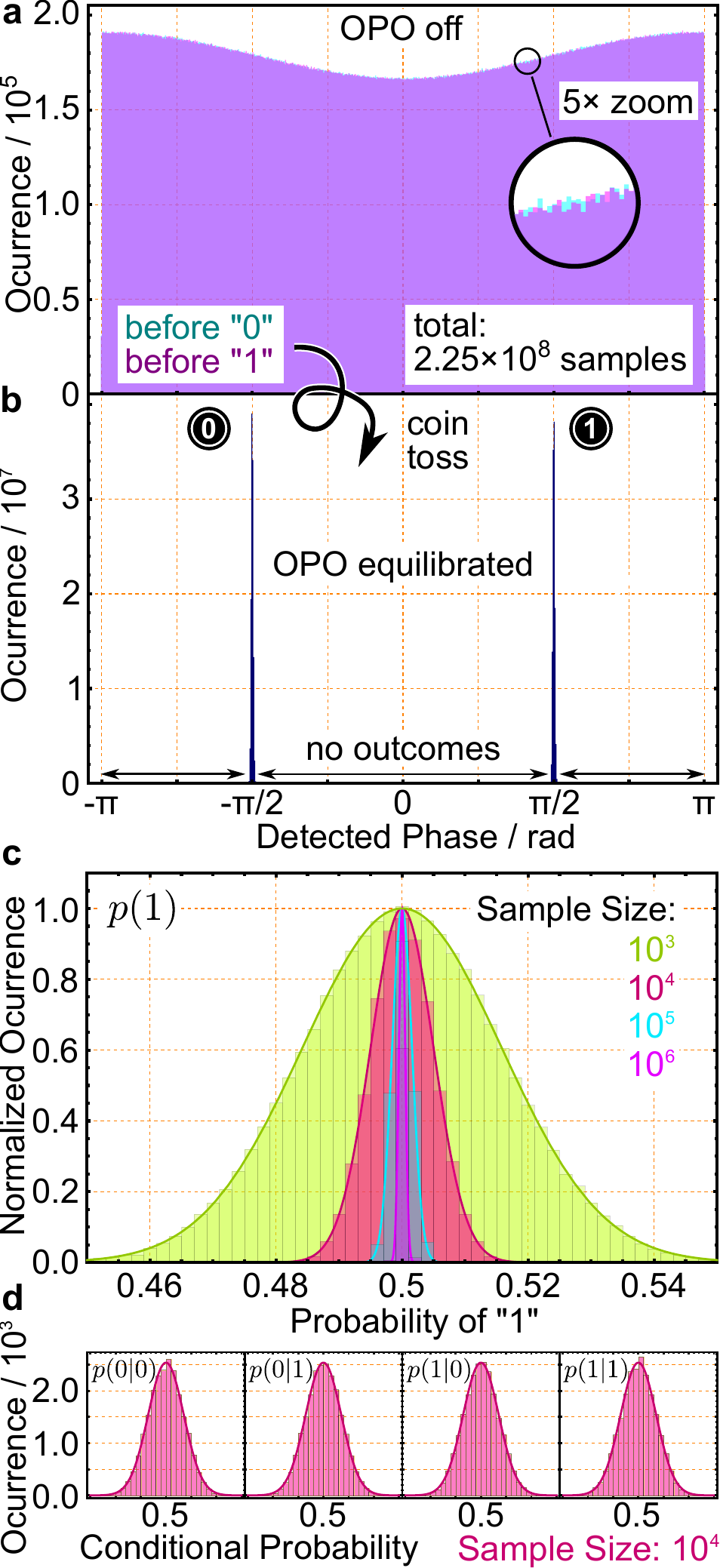}
  \caption{\textbf{Analysis of the raw bits.} a) Measurement outcomes when the OPO is off. Essentially all different phases are randomly measured, with a small bias. b) Measurement outcomes, after the OPO is equilibrated in the P2-state. c) Probability to find a \done \ as the outcome for different sample sizes, $N$. To note: The solid curve is the predicted result and not a fit. d) Conditional probability for the different options of tuple outcomes. Range spans from $p_{\mathrm{cond}}=$ 0.47 to 0.53. These probabilities are the relevant key figures for the entropy estimation below. Total sample size for all of the above: 2.25 $\times$ 10$^8$ measurements.}
  \label{fig:fig03}
\end{figure}

When the pump frequency (40.9~MHz in this case) is electronically divided by two, the pulse-train in the P2-state has a defined phase against this derived reference signal. When the OPO is turned on, this phase may be either in phase, or, with 50\% probability, out of phase. This phase difference of $\pi$ can be unambiguously measured with various demodulation techniques. A simple and convenient way is the relative multiplication between the detected signal and the reference. A simple commercial solution is the detection with a lock-in amplifier, which allows for a direct access to the relative phase, $\varphi$. Here, a Zurich Instruments lock-in amplifier is used (UHFLI). The measurement time to determine the phase amounts to 1~$\upmu$s.

For random number generation, the OPO is turned on and off by an optical chopper, which is installed such that it can inhibit the cavity oscillation. Fig.~\ref{fig:fig02}c shows the sequence of generating one single bit in the generator: The measured signal (red) is measured versus the reference signal (REF), which corresponds to half of the repetition rate of the pump laser ($f_{\mathrm{rep}}$). 
This measurement is performed twice in one chopper cycle: When the OPO is off -- as the control signal -- and when the OPO is in the P2-state -- as the signal of the running oscillator, the tossed and landed coin. The control measurement is performed to verify that two subsequent measurements do not carry spurious information from one to the next outcome. A sequence of four consecutive measurements in the on-state is depicted in Fig.~\ref{fig:fig02}d. H and L denote the two alternating, high and low pulse energy outputs of the OPO in the P2-state, respectively.

The measurement outcome is saved by a Matlab (Matlab Inc.) script into a comprehensive set of data, which saves all measured phases. These can be either analyzed as direct phases, or alternatively processed as bit outcomes.

The measured phase of the oscillating OPO exhibits essentially two measurement outcomes: $-\pi/2$ and $\pi/2$.  By means of a simple threshold the measurements are selected into a binary outcome. Values above zero phase are associated with the outcome \done, whereas values below zero are assigned a value of \dzero. Equally, these outcomes are the two possible stable configurations of the P2-state, $\text{L}\mathbf{H}\text{L}\mathbf{H}\dots $ (\dzero) or $ \text{H}\mathbf{L}\text{H}\mathbf{L}\dots$ (\done), where the order is fixed by the reference signal, at half of the pump frequency (see Fig.~\ref{fig:fig02}d). In the description above, a bold character denotes that the pulse from the OPO is not coinciding with the reference pulse train. This corresponds to a (red) colored character in Fig.~\ref{fig:fig01} or \ref{fig:fig02}. The measurement results were plotted in a histogram, and exhibit a very narrow distribution around the estimated value (see Fig.~\ref{fig:fig03}b).

\section{Origin of Randomness}

It is well established in the literature that the randomness element in the transient process of a starting OPO originates from quantum effects. These include vacuum fluctuations in the gain element as well as cavity losses~\cite{herrero-collantes_rmp_2017,wang_coherent_2013,marandi_oe_2012,drummond_critical_2002,nabors_coherence_1990,harris_observation_1967,louisell_quantum_1961}. The primary quantum process in the build-up of the oscillation is the generation of single photons in a spontaneous down conversion process caused by pumping the non-linear gain crystal~\cite{marandi_oe_2012,louisell_quantum_1961,harris_observation_1967}. The exact contribution of these processes to the formation of the P2-state is currently under investigation. In the context of randomness generation, it is important to note that the period doubling attractor is in particular not a chaotic attractor~\cite{uchida_fast_2008,kanter_np_2010}. This is despite the fact that period doubling and chaos might occur in one and the same nonlinear system, as outlined in detail in the supplementary material~\cite{steinle_supplemental_2017}. 

The independence of the primary randomness process against small fluctuations of the pump power is a crucial feature. In order to demonstrate this peculiarity, we have performed numerical pulse propagation simulations (RP Pro Pulse from RP Photonics) of the transient process with an artificially fixed additional seed. These show that a relative intensity change of more than $\pm 1 \% $ is required to induce a phase change by $\pi$ in the measured outcome. However, the measured relative intensity noise~\cite{steinle_ultra-stable_2016} integrated from 10~kHz to 20~MHz amounts to $\pm 0.0215 \% $ and is thus approximately a factor of $50$ too low to be the relevant driver of the randomness generation. 
 
Moreover, the independence of subsequent measurement outcomes is important, as discussed on the observed bits below. Therefore, the inter-bit waiting time was reduced in an additional experiment by a factor of 1000. This was performed with the OPO operated in an extended cavity configuration, such that four independent pulses oscillate simultaneously in the cavity. A subsequent measurement reads four bits within a single chopper cycle. This reduces the relevant timescale for the comparison of successive bits from 100~$\upmu$s to 100~ns and thus eliminates the contribution of mechanical vibrations, chopper jitter, thermal effects, and pump intensity noise. Nevertheless, we measure alternating bits, which would not be the case if any of the above technical effects would cause the randomness (see supplementary material~\cite{steinle_supplemental_2017}). These investigations indicate that quantum effects are a significant source of randomness in our system.

In order to further quantify the randomness this process produces, we analyze the measured phase and its binary representation for a large set of outcomes in the next section.

\section{From raw-bits to final bits}

The first analysis of the acquired data involves the measured phase $\varphi$ of the OPO in its off state. Fig.~\ref{fig:fig03}a shows a histogram of the raw phase output of the lock-in, right before each measurement of the running OPO. The data itself is divided into outcomes preceding an outcome of \dzero, and outcomes preceding an outcome of \done. Evidently, both datasets are very similar, and do not show any particular preference for subsequent outcomes. The small bias (wavy curve) is based on spurious signals reaching the lock-in amplifier and is symmetric for both phase outcomes.

After the transient time has passed, a second measurement determines the final state (=OPO on). As above, this is analyzed by the lock-in amplifier, resulting in a histogram of events. Both possible outcomes are centered around $-\pi/2$ and $\pi/2$, respectively. Their distribution is determined by experimental uncertainty to determine the phase. This results from spurious phase information, spontaneous down-conversion in the crystal, the sampling and measurement time, and residual (phase) noise in the signal. The width of the determined outcomes ($1 \sigma$) amounts to 0.0023~rad. In other words, the outcomes are separated by more than 400 standard deviations -- excluding the possibility that the two outcomes are confused. Such ambiguity-free measurements cannot be achieved in generators which are based on photon counting due to e.g.\ dark counts~\cite{stefanov_jmo_2000,svozil_pra_2009,oberreiter_light_2016}.

In the course of approximately one day a number of 2$\times$2.25$\times$10$^8$ measurements were performed. We now analyze a possible bias or imbalance of the experimental outcomes, caused for example by technical noise~\cite{mitchell_pra_2015}. 
This noise would produce additional measurement outcomes, which in information theoretical terms add up to the randomness in the transient process of the generator. For the analysis, the bit stream is divided into sub-strings of length $N$ and the experimental probability of the outcome \done \ is determined. The distribution is centered around 0.5, independently of the sample size $N$. The analysis reconfirms the width of the distribution as $\sigma_{\mathrm{single}}=\sqrt{N p (1-p)}/N$. 
Please note that the data is not fitted but the theoretical curve is depicted along with the measured data.

The balance of the measurement outcomes is only one indication of a well balanced coin toss. Another important measure is the \emph{conditional} probability which signifies whether subsequent outcomes contain some form of memory of the prior state of the oscillator. For this, a first indication is given by the analysis of Fig.~\ref{fig:fig03}a and b -- still, this does not prove the independence of the outcomes of subsequent measurements in the equilibrated OPO. The conditional probability of obtaining the result \done \ after a preceding result \dzero \ is denoted as $p(\done|\dzero)$, reading as the probability of one conditioned on zero. This is defined as $p(x|y)=p(x \land y)/p(y)$, and is depicted in Fig.~\ref{fig:fig03}c, along with the theoretical prediction of its distribution $\sigma_{\mathrm{cond}}=1/\sqrt{2N}$. An auto-correlation analysis, which also accounts for higher order bit-to-bit correlations is given in the supplementary material~\cite{steinle_supplemental_2017}. Again, the expected behavior is reconfirmed and no memory in the system is evident.

Very common is the use of so-called \emph{random number tests}. The tests \emph{ent}, the \emph{NIST test suite}~\cite{rukhin_r_2010}, the \emph{die-harder} suite, or the most comprehensive \emph{TestU01} suite~\cite{lecuyer_atoms_2007} are commonly known. Many people still believe that such tests are able to show whether a bit-string is random or not. But they can only deliver the proof that no substantial flaw occurred in the implementation of a random bit generator. Moreover, most of these tests are based on algorithmic information theory and are designed to test algorithmically generated pseudo-random numbers rather than random numbers generated by physical processes~\cite{calude__2010}. Therefore, the statement that a certain bit-string passes all tests does not prove the random nature of the input. Non-random and predictable numbers, such as the binary expansion of $\pi$, pass all these tests flawlessly. As expected, our presented generator passes all these tests, and a sample output for the NIST suite is presented in the supplementary material~\cite{steinle_supplemental_2017}.

A subset of the described random number tests is the analysis of different bit-patterns and their occurrence in the data set. This approach has been examined in early discussions on random number testing~\cite{knuth__1968}. Nowadays, other authors suggest the use of information theoretic language for random number testing~\cite{calude__2010}. In this context the \emph{coin tossing constants} by \textsc{W.\ Feller}, which are closely related to the generalized \textsc{Fibonacci} numbers~\cite{finkelstein_fq_1978}, describe the asymptotic probability $p(n,k)$ of the event that a sequence with the length $k$ of \done \ or \dzero \ does \textbf{not} occur in a sequence of $n$ tosses of a fair coin. \textsc{Feller}'s constants have the property
\begin{equation} 
\lim_{n\rightarrow \infty} p(n,k) \alpha_k^{n+1}=\beta_k \, .
\end{equation} 

\noindent
The given Table 1 displays the analysis of sub-strings of length $N=$ 400 bits of the generator. This small number is chosen to have non-vanishing values for the probabilities associated with higher order parameters ($k>5$). The experimentally determined value is given in the third column, and the relative deviation of the order of 10$^{-4}$ corresponds to the square root (=shot noise) of 2.25$\times$10$^8$ recorded bits. The computed values of the coin tossing constants match very well to the assumed behavior of the supplied random bit sequence.

The coin tossing constants analyze higher orders of tuples than the conditional probability and are therefore similar in this respect to a mathematical \textsc{Borel}-\emph{normality} test~\cite{knuth__1968}, which analyzes the lexicographical occurrence of all possible binary strings. Such a test was implemented by \textsc{C.\ Calude} for testing a number of (hardware-based) randomness generators~\cite{calude_pra_2010}.

The above analysis on the probability of subsequent sets of measurement outcomes underlines the behavior of an ideal coin toss. An interesting effect occurs when we process the measurement outcomes by pairing each bit with exactly one neighboring bit, without allowing any overlaps of the tuples -- unlike as before. Although we find all tuple permutations (\dzero\dzero, \dzero\done, \done\dzero, \done\done) to be equally probable, the waiting time, which is the ``distance'' between two equivalent outcomes is different between the bit-changing (\dzero\done, \done\dzero) and bit-equivalent outcomes (\dzero\dzero, \done\done). For the tuples including a bit-flip, the predicted waiting time is 4 consecutive tosses. On the other hand a double sequence of \dzero\dzero, or \done\done, has a predicted waiting time of 6 consecutive tosses. This is verified with the present set of data and we determine values of 3.99976 and 5.99784 respectively. Again, the relative uncertainty of approximately $10^{-4}$ corresponds to the length of the data-set; it proves that there is no further memory storage in the measurement outcomes and reconfirms the predicted behavior.

In summary, we conclude that the measured raw bits of the presented all-optical randomness generator using a nonlinear feedback OPO in the P2-state do not differ by any measurable means from the ones of a perfect Bernoulli trial. This is indicated by the independence of consecutive measurement outcomes, the balance between the two probabilities, and further tests, which resemble the expected outcomes of a perfect coin toss. Subsequently, the required post-processing can be reduced to a minimum. Such a post-processing would generally be required for any physical implementation of a fair (perfect) coin-toss due to finite size effects. We now turn to the entropy analysis of the raw-bit stream.


\begin{table}[hb]
\label{tab:tab01}
\begin{tabular}{|l|l|l|l|}
\hline
$k$&$\alpha_{\mathrm{ideal}}$&$\alpha_{\mathrm{extracted}}$&Relative change\\ \hline
2&1.23606798&--&--\\
3&1.08737803&--&--\\
4&1.03758013&1.03676354 & 7.87010735$\times$10$^{-4}$\\
5&1.01732078&1.01731406 & 6.61125775$\times$10$^{-6}$\\
6&1.00827652&1.00827933& -2.78877013$\times$10$^{-6}$\\
7&1.00403411&1.00403701& -2.88459780$\times$10$^{-6}$\\
8&1.00198836&1.00198588& 2.47363715$\times$10$^{-6}$\\
9&1.00098624&1.00098584& 4.01117501$\times$10$^{-7}$\\
10&1.00049092&1.00049182& -8.99357769$\times$10$^{-7}$\\
11&1.00024486&1.00024624& -1.38152744$\times$10$^{-6}$\\
12&1.00012226&1.00012358& -1.31441456$\times$10$^{-6}$\\
13&1.00006109&1.00006163& -5.40416736$\times$10$^{-7}$\\
14&1.00003053&1.00003025& 2.79986856$\times$10$^{-7}$\\
15&1.00001526&1.00001522& 4.33916550$\times$10$^{-8}$\\ \hline
\end{tabular}
\caption{\textbf{Feller's coin tossing constants.} The constants are related to the probability that a certain sequence of \done s does not occur in a set of random bits. Here, the sample size is $N=400$. The ideal value of the coin tossing constant $\alpha$ is compared to the values extracted from our experimental data. Relative change is calculated as $(\alpha_{\mathrm{ideal}}-\alpha_{\mathrm{extracted}})/\alpha_{\mathrm{ideal}}$ . The relative uncertainty is given by the finite length of the acquired data set.}
\end{table}

\begin{figure*}[ht!]
  \includegraphics[width=\textwidth]{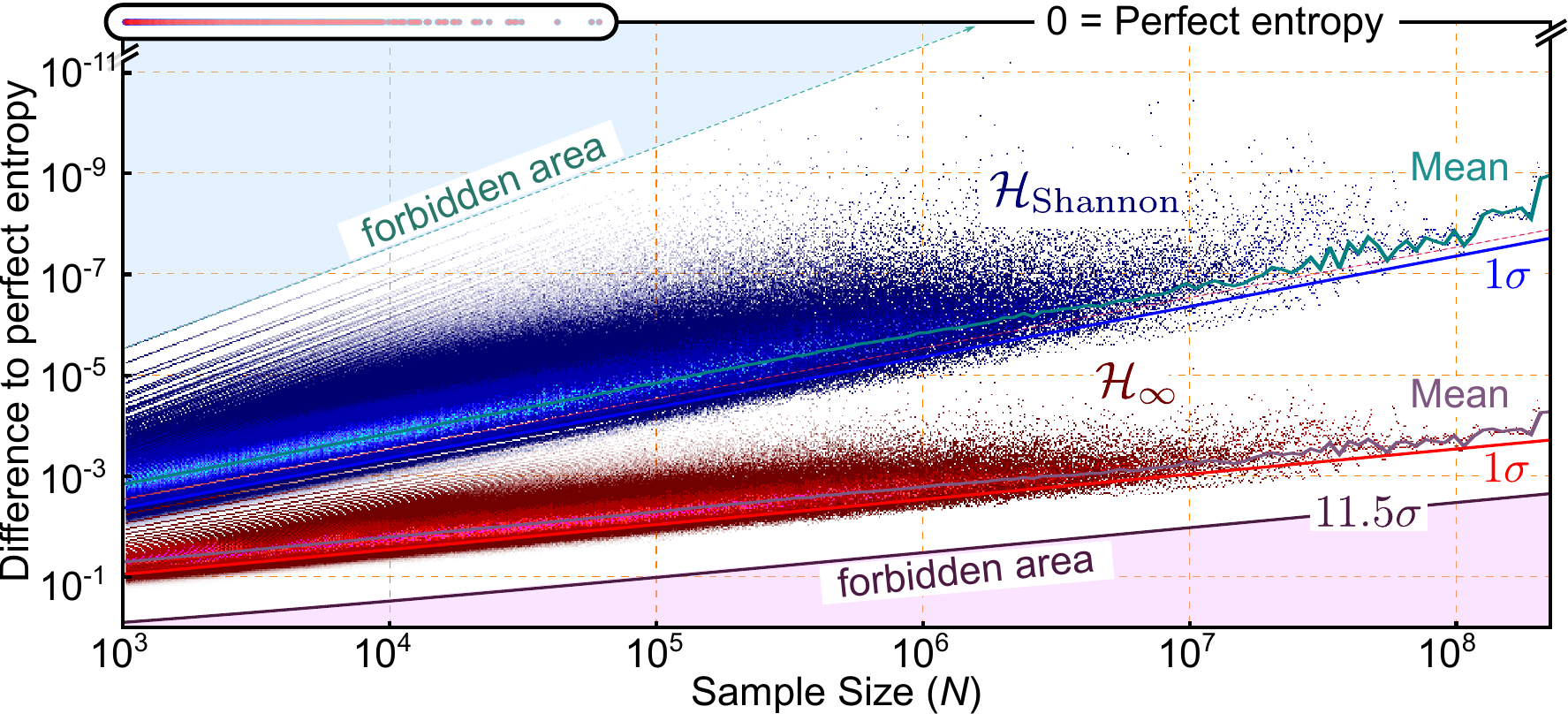}
  \caption{\textbf{Final entropy in the generated bit stream.} The difference of the entropy ($\mathcal{H}$) to unity is shown for the Shannon-entropy (blue) and min-entropy (red) against the sample size ($N$). A higher density and brighter color of the points obtained from experimental data signifies more outcomes of a certain value. The best possible case is a difference of 0, as displayed after a cutoff of the logarithmic scale on top of the graph. For $ N<10^5 $, sequences possessing this ``perfect'' entropy are still observed, shown as separate points. If a specific single bit is flipped, the entropy is reduced below unity. Subsequently, points in the graph which do not exhibit unity entropy cannot be higher than a certain limit (dashed curve). This forms a forbidden area to the top with no mathematically possible outcome. The solid straight line at the bottom indicates the conservative bound. These bounds are obtained a priori by an error propagation on the entropy of a fair coin as outlined in the supplementary material~\cite{steinle_supplemental_2017}. Red: one $\sigma$ deviation from the expected min-entropy. Purple: Assuming an outlier probability of 2$^{-100}$. As expected, no values are found below this line.}
  \label{fig:fig04}
\end{figure*}

\section{Entropy estimation}

While all above measures suggest that the raw-bits are usable as a perfect source of random bits, we have ignored an important information theoretical measure of the output of the experimental apparatus so far: The generated entropy. As outlined below, the crucial quality figure for a randomness generator is the achievable entropy per output bit. Ideally, each bit has the perfect entropy of unity, which means that each generated bit can be used as an independent optical coin-toss and resembles the output of a fair coin. But when a finite fraction of bits is analyzed, this can only be proven if all \done s and \dzero s are equally balanced. Intrinsically, there might be an unwanted (but statistically allowed!) bias. In this case the determined entropy will be lower than one. Due to the finite length, this is most likely the case for the presented data-set. A first naive approach to calculate the entropy analyzes the balance of the bit-stream, and is given by the unconditional Shannon entropy which is defined as
\begin{eqnarray}
\mathcal{H}_{\mathrm{Sh}} = \sum_{y} p(y) I(p(y)) = - \sum_{y} p(y) \log_2 p(y) \ .
\end{eqnarray}

\noindent
Where $ p(y) $ is the single probability of obtaining $ \dzero $ or $ \done $ in the full bit sequence, respectively. This, however, does not consider any dependence or memory effects in the measurement outcomes, where for example an alternating sequence \done\dzero\done\dzero\done\dzero\ldots would result in the same entropy as a fully random, i.e.\ totally unordered sequence. Therefore, the \emph{conditional} entropy is considered, accounting for the memory (or the absence thereof) in the system. This is defined as 

\begin{eqnarray}
\nonumber 
\mathcal{H}_{\mathrm{Sh}}(X|Y)&=&\sum_y p(y) \mathcal{H}_{\mathrm{Sh}}(X|Y=y)\\
&=&-\sum_y p(y) \sum_x p(x|y) \log_2{p(x|y)} \ .
\end{eqnarray}
\noindent
We refer to the supplemental material~\cite{steinle_supplemental_2017} for details of our calculation of the conditional entropy, but mention for clarity that the events y and x are defined as ``the $ i $-th bit is $ \dzero $($ \done $)'' and ``the $ (i+1) $-th bit is $ \dzero $($ \done $)''. Uppercase Y and X are the unified sets of events on all bits. Thus, our notion of entropy is linked to the frequency analysis of output data, but can also be estimated a priori.
\noindent
Unlike the Shannon entropy, the min-entropy (denoted as $\mathcal{H}_{\mathrm{\infty}}$) is the most conservative bound for the usable entropy of a randomness generator. It maximizes the (conditional) probability $p(x|y)$ against $x$. This imbalance and maximizing effect can be seen in Fig.~\ref{fig:fig03}c and d. It becomes evident, that for a larger sample size $N$, the width of the distribution shrinks and the amount of entropy is commonly larger.
The min-entropy is defined as
\begin{equation}
\mathcal{H}_{\mathrm{\infty}}(X|Y) =  -\log_2 \left[ \sum_y p(y) \max_x \left\lbrace p(x|y)  \right\rbrace \right] \ .
\end{equation}

\noindent
The above entropy definitions can be straight-forwardly computed for an experimentally generated data-set. This will result in a scalar entropy value, which still has to be interpreted; for a good generator the resulting number will be usually close to one. How ``perfect'' the entropy is and how close it reaches to one, depends on three factors: a) the quality of the generator, b) the size of the analyzed bit stream (here denoted as $N$ for the number of analyzed bits), and, c) which particular data set is analyzed. Conclusively, it is very unlikely to achieve an entropy of unity when the entropy for a finite bit string is computed. This even holds for a fair coin. In the following, we perform an analysis of the generator's outcome and compute if the entropy matches the predicted value. 

Fig.~\ref{fig:fig04} shows the calculated Shannon and min-entropy for the presented data-set against the sample size, $N$. To note that this graph shows the deviation against perfect entropy on a logarithmic scale. For a \emph{smaller} sample size $N$ (left-hand side), a larger number of samples exist, and more points are depicted. As mentioned before, with a \emph{larger} sample size, $N$, the entropy approaches unity. 
The conditional Shannon-entropy scales linearly with $N$, whereas the min-entropy is proportional to $\sqrt{N}$. The value and the distribution of the min-entropy is significantly smaller than for the Shannon-entropy, since the conditional probability is maximized. Fig.~\ref{fig:fig04} also shows entropy bounds which are obtained a priori. These include the second highest possible value of the entropy for a certain sample size, besides the ideal case of perfect entropy. This is the highest value which can occur, when a minimally entropy-changing single bit-flip is present in a dataset of length $N$. These curves scale quadratic against the mean slope behavior which was introduced above. Therefore, the mean value for the conditional Shannon-entropy forms a parallel line to the highest min-entropy, where one bit-flip is present.

The min-entropy is a conservative bound and selects the maximal conditional probability in a set of random bits. If a perfect random string is infinitely long, every possible occurrence will show up in a subset of this sequence. Then, in contradiction to the description above, a set of calculated entropies would eventually be very small since a very long sequence of seemingly non-random bits can occur (e.g.\ such as \done\done\done\done\done\done\done\ldots). For these cases, the calculated entropy may be reduced to zero. For realistic considerations it is therefore important to exclude for instance such infinitesimally likely events of all bits of a long sequence being \done. Such a calculation of the occurrence of a certain set of equivalent outcomes of a generator was presented in the calculation of the coin tossing constants above (Table 1). Additionally, a possible error bound for randomness extraction was introduced by \textsc{M. Troyer} and \textsc{R.\ Renner}~\cite{troyer_ir_2012} as $1/2^{100} \approx 1/10^{30}$. Such bounds are also described to guarantee an ``$\epsilon$-randomness''~\cite{mitchell_pra_2015}. The proposed bound of $1/2^{100}$ ensures that one million generators do not have the option to exhibit the same outcome (i.e.\ a so-called \emph{collision} of two generators) in the age of the universe. In the case of Gaussian distributed events this corresponds to approximately 11.5 standard deviations from the center of the distribution. Fig.~\ref{fig:fig04} shows this bound as the lowest curve, obtained a priori by an error propagation on the entropy of a fair coin as outlined in the supplementary material~\cite{steinle_supplemental_2017}.
As suggested from the raw bit analysis, no selected sub-set of the bits falls below this line -- this suggests that the model of a perfect coin toss  seems to be appropriate for the introduced generator.

For our presented sample size of 2.25$\times$10$^8$ the conditional min-entropy per bit can be estimated as 99.95\%. This can be simply read from Fig.~\ref{fig:fig04} on the right hand side. This is, of course, solely limited by the finite sample volume. 
The most conservative bound (11.5 $ \sigma $) of the entropy difference to unity is approximately one order of magnitude different, and the entropy amounts to 99.5\%. 

With the raw bits, as discussed above, but also by merit of the calculated entropies, we are able to prove that the recorded bit-stream does not differ by any measurable means from a perfect coin toss. Each emitted bit can therefore be used as a random bit. No further randomness extraction has to be considered, when a large enough bit string is used. Of course, we are only able to prove this assumption bound to the size of the recorded bit string.

\section{Conclusion \& Outlook}

An unbiased all-optical coin toss has been presented. It is based on the bi-stable outcome of an optical parametric oscillator with nonlinear fiber feedback, operating in the P2-state. The detection scheme relies on phase detection versus an external reference pulse. This implementation is substantially simpler than prior published experiments~\cite{marandi_oe_2012,marandi_oe_2012a,okawachi_ol_2016}, since it does not require degenerate operation of the OPO. The disadvantage of degenerate operation is that it necessitates either an actively interferometrically stabilized resonator to fix the relative optical phases of the signal and idler frequency combs to the pump frequency comb, or a ``shaker'' using a ``dither and lock'' algorithm that periodically varies the cavity length to generate an error signal for the stabilization. This introduces noise to the system which can be avoided by a non-degenerate operation.

The implemented detection scheme, based on period doubling, is ambiguity free, i.e.\ has only two possible outcomes, separated by more than 400 standard deviations, which can be interpreted as zeros and ones of a random bit sequence. This uniquely decouples the fundamental randomness process from the detection principle. While the detection here is based on a lock-in amplifier, more simple schemes can be developed. A demodulator or a radio-frequency mixer and a comparator will reduce the implementation costs, and emit  the random sequence directly into an e.g.\ TTL level output.

One limitation is given by the sample rate of the chopper, which is limited to 10~kHz in the presented design. This sample rate is ultimately limited by the transient process until the OPO is in a stable state and the required time for phase detection. The measurement time to determine the phase amounts to 1~$\upmu$s with the current detection system. This may be shortened in future experiments by a factor of 10. Accordingly, a faster chopper can be installed as well. As evident in Fig.~\ref{fig:fig02}c, we estimate the time for equilibration to approximately 300~ns and the ambiguity-free detection of the phase state to two to three cycles, amounting to 100-150~ns. With the described OPO, and by introducing a faster chopper, a random bit rate above 1~MHz can be reached. An even further speed-up can be implemented with a higher repetition rate of the pump laser. For such changes, OPOs reaching the GHz range are reported~\cite{roth_iptl_2004}. As a side-effect, this would result in a much more compact design for the entire experimental configuration.
Building a more compact randomness generator could further be realized by implementing the introduced principle with state-of-the-art technology on a photonic chip~\cite{niehusmann_ol_2004,kuyken_oe_2013,abellan_optica_2016}.

The full quantum mechanical description of the open quantum system, specifically in the P2-state, remains to be addressed in future work. Commonly, the process of a bi-stable outcome of an OPO is described as a quantum process~\cite{herrero-collantes_rmp_2017,wang_coherent_2013,marandi_oe_2012,drummond_critical_2002,nabors_coherence_1990,harris_observation_1967,louisell_quantum_1961}, growing from quantum mechanical vacuum fluctuations. A careful analysis on the transient process, which may also introduce a fiber-optic electro-optic modulator instead of a chopper, along with more research on the power dependence will likely characterize this process and the P2-state in further detail. A deeper understanding of the underlying physics might lead to faster phase detection and larger random bit rates, and even to future implementations in quantum information processing and quantum simulation~\cite{inagaki_s_2016}.

\vspace{3mm}
\begin{acknowledgments}
We acknowledge the support for the 3D-rendering by Ingmar Jakobi for Fig.~\ref{fig:fig01}. We further acknowledge the funding from the MPG, the SFB project CO.CO.MAT/TR21, ERC (Complexplas), the BMBF, the Eisele Foundation, the project Q.COM, and SMel. 
\end{acknowledgments}


\end{document}